\documentclass[12pt]{article}
\usepackage{amssymb}
\usepackage{amsmath}
\usepackage{hyperref}
\usepackage{graphics}
\usepackage{pstricks}
\usepackage{mdframed}
\usepackage{graphicx} 

\begin{document}
\title{\bf Holographic Aspects of a Higher Curvature Massive Gravity}
\author{Shahrokh Parvizi$^1$\thanks{Corresponding author: Email: parvizi@modares.ac.ir}  \hspace{2mm} and
      Mehdi Sadeghi$^{2}$\thanks{Email: mehdi.sadeghi@abru.ac.ir}\hspace{2mm}\\
		{\small {\em $^1$Department of Physics, School of Sciences,}}\\
        {\small {\em Tarbiat Modares University, P.O.Box 14155-4838, Tehran, Iran}}\\
       {\small {\em  $^2$Department of Physics, School of Sciences,}}\\
 {\small {\em  Ayatollah Boroujerdi University, Boroujerd, Iran}}\\
       }
\date{\today}
\maketitle

\abstract{We study the holographic dual of a massive gravity with Gauss-Bonnet and cubic quasi-topological higher curvature terms. Firstly, we find the energy-momentum two point function of the 4-dimensional boundary theory where the massive term breaks the conformal symmetry as expected. An $a$-theorem is introduced based on the null energy condition. Then we focus on a black brane solution in this background and derive the ratio of shear viscosity to entropy density for the dual theory. It is worth mentioning that the concept of viscosity as a transport coefficient is obscure in a nontranslational invariant theory as in our case. So although we use the Green-Kubo's formula to derive it, we rather call it the rate of entropy production per the Planckian time due to a strain. Results smoothly cover the massless limit.  }\\



\section{Introduction} \label{intro}

\indent For two decades, the AdS/CFT correspondence \cite{Ref7}-\cite{Gubser:1998bc} has been at the center of attention in theoretical physics. It not only provides tools for performing calculations in strong coupling limit of field theories and condense matter phenomena, which otherwise undoubtedly was horrible if not impossible, but also opens new windows to understand different aspects of field theories and gravities as well. Much has been done for the Einstein gravity in the AdS bulk and investigated its CFT dual on the boundary. In the early of the AdS/CFT, the central charges of the boundary theory were found by the holography \cite{Henningson:1998gx}. This was an important success which among others  encouraged people to develop the duality for more complicated and realistic theories. In 4 dimensions, it is well known that CFT's with Einstein gravity dual have two equal central charges $a$ and $c$ \cite{Henningson:1998gx,Henningson:1998ey}. This means that the dual Einstein gravity gives information about a very special class of CFT's. To distinguish between central charges and explore more general conformal theories, one may add higher curvature (or higher derivative) gravities, which amongst them, the Gauss-Bonnet  gravity serves as a simple model to study the duality. It has important features as its equation of motion includes only the second order derivatives, admits exact black hole solution and the corresponding dual theory would be a CFT with two distinguished central charges $a\neq c$ \cite{Nojiri:1999mh}. It is possible to keep these advantages and add cubic curvature terms, the so called quasi-topological gravity which of course doesn't generally admit a second order differential equation except for the AdS background which is in our interest \cite{Oliva:2010eb,Myers:2010ru,Myers:2010jv}. The combination of Gauss-Bonnet and cubic quasi-topological gravity has not yet any stringy derivation, however as a toy model is rich enough to study different aspects of the dual conformal theory \cite{Myers:2010jv}.  

On the other hand, several studies have been performed for decades to generalize the graviton field to a massive one with different motivations from theoretical curiosity to phenomenological model buildings (for a recent review see \cite{deRham:2014zqa}). Indeed, the problem of giving mass to the gravity is not an obvious one and was a challenge for several years. The first attempt was by Fierz and Pauli \cite{Fierz:1939ix}  proposing a linear massive model. Unfortunately, that doesn't reduce to GR in the zero mass limit. Generalization to a nonlinear model in \cite{Isham:1971gm} was stopped by Boulware and Deser (BD) when they showed that this suffered from ghosts \cite{Boulware:1973my}. Finally in recent years, the BD ghost problem was resolved in \cite{deRham:2010ik,deRham:2010kj,Hassan:2011hr} by a nonlinear massive gravity. This theory provides a fixed reference metric on which the massive gravity propagates. This breaks the general covariance with applications in holographic models with momentum dissipation \cite{Vegh:2013sk,Davison:2013jba}. Then it was extended to include dynamics of this reference metric in the context of theories now known as bi-gravities \cite{Hassan:2011zd,Hinterbichler:2012cn,Nomura:2012xr} and  higher dimensional massive graviton term is discussed in \cite{Do:2016uef}. Many recent works include this massive gravity in the higher curvature gravities especially the Gauss-Bonnet ({\it e.g.}, \cite{Sadeghi:2015vaa,Hendi:2015pda,Hendi:2016yof}) and various features mostly in the gravity side and some in the dual theories are derived. 

Here our aim is to tackle the theory including  Gauss-Bonnet cubic quasi-topological massive gravity. The important point about this combination is that while it has the rich structure of higher curvature theories, the presence of a mass scale breaks the conformal symmetry on the boundary. Indeed, the quantum field theory dual to the Gauss-Bonnet theory is not well-known and there are doubts if it even exists \cite{Camanho:2014apa,Cheung:2016wjt}. This might be the case for a dual to the massive gravity. However, the Gauss-Bonnet gravity is widely discussed in holographic literature at least as a toy model or theoretical laboratory to study general aspects of quantum field theories or CFT's. In this direction, one may consider the addition of massive gravity as a relevant perturbation of conformal symmetry. This may shed light on the boundary theory, if any, away from the fixed point. By the way, we emphasize that this set up should be considered as a toy model to study the holography.

In this regards, we assume a boundary theory dual to our bulk model in section 2. Firstly we derive the two-point function of the boundary energy-momentum tensor and show that it includes a massive operator on the boundary. 
We then look for an $a$-theorem which is originally based on renormalization group flows \cite{Zamolodchikov:1986gt,Cardy:1988cwa,Komargodski:2011vj,Komargodski:2011xv} and indicates the truncation of degrees of freedom when going toward an IR fixed point. In the context of AdS/CFT correspondence, $a$-theorems are introduced by considering a generic background which asymptotes to AdS space \cite{Girardello:1998pd,Freedman:1999gp,Oliva:2010eb,Myers:2010tj,Liu:2010xc,Liu:2011iia}. Then the $a$-function approaches the $a$ charge (not the $c$) in the AdS limit. The function should be monotonically decreasing along the RG flow. This can be achieved by the null energy condition. In our case, we show that the same null energy condition as the massless theory gives the correct monotonic $a$-function.

In section 3, we introduce an exact black brane solution in this background and derive its temperature and entropy then using the standard holographic methods to find the viscosity to entropy ratio. It is worth noting that in a theory where the translational symmetry is broken, e. g. by a mass term as in our case, the viscosity can not be interpreted as a hydrodynamic transport coefficient. Instead it can be considered as the rate of entropy production per the Planckian time due to a strain \cite{Hartnoll:2016tri} and can be derived from the Kubo formula (see Eq. (\ref{Kubo}))\footnote{There are some articles, e.g. \cite{Davison:2014lua,Alberte:2016xja}, that take this later quantity to be same as the viscosity as a transport coefficient. However, we take these two quantities be different.}. On the other hand, \cite{Burikham:2016roo} introduces a shear viscosity from hydrodynamic constitutive relations. Both quantities approach the same viscosity in the massless limit. In our case we deal with the first one, but sometimes we may call it simply `viscosity' rather than the rate of entropy production. 

In the Einstein gravity with any matter content, the ratio $\eta/s$ is found to be $1/4\pi$ \cite{Buchel:2003tz}. It was proposed by the KSS that this is a lower bound for relativistic quantum theories \cite{Ref27}. However, in the higher curvature gravities this bound is violated \cite{Brigante:2007nu}. For the massive gravity, the `viscosity' was calculated in \cite{Sadeghi:2015vaa} and \cite{Hartnoll:2016tri} as a deformation of the metric component $\delta g_{\mu\nu}$ and shown that the naive bounds on $\eta/s$ are violated. However, \cite{Hartnoll:2016tri} has different interpretation for the bound and argues that it is expected to exist due to a basic quantum mechanical uncertainty.
Here we apply this interpretation and take $\eta$ as the rate of entropy production and extend the calculation to include higher curvature theories with mass term. The result may lower the `viscosity' to entropy ratio more than before. 

In section 4, we discuss on some bounds on parameters space. We consider the unitarity and causality bounds on the boundary theory.

\section{Quasi-Topological Massive Gravity and Holography}
\label{sec-conformal}

\indent Let us start with a quasi-topological Gauss-Bonnet massive gravity in 5-dimensions with a negative cosmological constant. The action is given by \cite{Myers:2010ru,Sadeghi:2015vaa},

\begin{align}\label{Action}
&S=\frac{1}{2\ell_p^3}\int d^{5}  x\sqrt{-g} \Bigg(R+\frac{12}{L^2}+\frac{\lambda L^2}{2}\mathcal{L}_{GB}+\frac{7}{8}L^4\mu \mathcal{L}_{3}+m^2\sum_{i=1}^4{c_{i}\mathcal{U}_i(g,f)}\Bigg)\\
&\mathcal{L}_{GB}=R^2-4R_{\mu \nu}R^{\mu \nu}+R_{\mu \nu \rho \sigma }R^{\mu \nu \rho \sigma}\nonumber\\
&\mathcal{L}_{3}=R_{\mu \nu}^{\,\,\,\,\,\,\,\,\rho \sigma}R_{\rho \sigma}^{\,\,\,\,\,\,\,\,\alpha \beta}R_{\alpha \beta}^{\,\,\,\,\,\,\,\mu \nu}+\frac{1}{14}(21R_{\mu \nu \rho \sigma}R^{\mu \nu \rho \sigma}R-120R_{\mu \nu \rho \sigma}R^{\mu \nu \rho}_{\,\,\,\,\,\,\,\,\,\,\alpha}R^{\sigma \alpha}\nonumber\\&+144R_{\mu \nu \rho \sigma}R^{\mu \rho}R^{\nu \sigma}+128R_{\mu}^{\,\,\,\,\nu}R_{\nu}^{\,\,\,\rho}R_{\rho}^{\,\,\,\,\mu}-108R_{\mu}^{\,\,\,\,\nu}R_{\nu}^{\,\,\,\mu}R+11R^3 )\nonumber
\end{align}
where $ R $ is the scalar curvature, $ L $ the cosmological constant scale, $f$ a fixed rank-2 symmetric tensor known as reference metric and $m$ is the mass parameter. $\mathcal{L}_{GB}$ and $\mathcal{L}_{3}$ are respectively  the Gauss-Bonnet and quasi-topology terms of gravity with $\lambda $ and $\mu$ their dimensionless couplings. A generalized version of the reference metric $ f_{\mu \nu} $ was proposed in {\cite{deRham:2010kj}} with the form $ f_{\mu \nu} = diag(0,0,c_0^2h_{ij})$ with $h_{ij}=\delta_{ij}/L^2 $. In (\ref{Action}), $ c_i $'s are constants and $ \mathcal{U}_i $ are symmetric polynomials of the eigenvalues of the $ 5\times5 $ matrix $ \mathcal{K}^{\mu}_{\nu}=\sqrt{g^{\mu \alpha}f_{\alpha \nu}} $,
 \begin{align}\label{7} 
	& \mathcal{U}_1=[\mathcal{K}]\nonumber\\
	& \mathcal{U}_2=[\mathcal{K}]^2-[\mathcal{K}^2]\nonumber\\
	&\mathcal{U}_3=[\mathcal{K}]^3-3[\mathcal{K}][\mathcal{K}^2]+2[\mathcal{K}^3]\nonumber\\
	& \mathcal{U}_4=[\mathcal{K}]^4-6[\mathcal{K}^2][\mathcal{K}]^2+8[\mathcal{K}^3][\mathcal{K}]+3[\mathcal{K}^2]^2-6[\mathcal{K}^4] 
\end{align}
The square root in $ \mathcal{K} $ means $ (\sqrt{A})^\mu_\nu(\sqrt{A})^\nu_\lambda=A^\mu_\lambda $ and the rectangular brackets represent traces {\cite{deRham:2010kj}}. 

The theory admits an asymptotic $AdS$ solution as follows
\begin{equation}\label{metric}
ds^{2} =-\frac{r^2N(r)^2}{L^2}f(r)dt^{2} +\frac{L^2dr^{2}}{r^2f(r)} +r^2h_{ij}dx^idx^j,
\end{equation}
where details of $N$ and $f$ functions are given in section (\ref{sec-blackbrane}). These details are not important for our purposes in this section and we restrict ourselves to the near boundary behavior where  $\lim_{r \to \infty}f(r)=f_\infty$ and  $\lim_{r \to \infty}N(r)^2f(r)=1$ which corresponds to the following $AdS$ background,
\begin{equation}\label{AdS}
ds^{2} =-\frac{r^2}{L^2}dt^{2} +\frac{L^2dr^{2}}{r^2f_\infty} +r^2h_{ij}dx^idx^j,
\end{equation}
with the radius of curvature 
\begin{equation}
	\tilde{L}^2=\frac{L^2}{f_\infty}
\end{equation}
$f_\infty$ is found from 
\begin{equation}\label{f-infinity}
	1-f_\infty+\lambda f_\infty^2+\mu f_\infty^3=0 
\end{equation}
A simple derivation is to take $r\rightarrow \infty$ limit of (\ref{f-equation}).

Then values of $ \mathcal{U}_i $ in (\ref{7}) are calculated as below,
\begin{align}\label{U1-U4} 
  & \mathcal{U}_1=\frac{3c_{0}}{r}, \,\,\,  \,\,\, \mathcal{U}_2=\frac{6c_0^2}{r^2},\,\,\,\,\mathcal{U}_3=\frac{6c_0^3}{r^3},\,\,\,\,\mathcal{U}_4=0 \nonumber\\
  &  \mathcal{U}\equiv  m^2\sum_{i=1}^4{c_{i}\mathcal{U}_i}=m^2\Big(\frac{3c_0c_1}{r}+\frac{6c_0^2c_2}{r^2}+\frac{6c_0^3c_3}{r^3}\Big)
   \end{align}

\subsection{Holographic Picture on the Boundary}
 Here we try to shed light on the dual boundary picture, if any, by the standard AdS/CFT prescription. Let's start with the theory on $AdS_{d+1}$ background and try to find the correlation functions on the boundary. Since we are dealing with pure gravity in the bulk, it is natural to look for energy momentum two-point function on the boundary. In a conformal field theory, the symmetry dictates the form of two-point function to be \cite{Gubser:1997se}
\begin{equation}\label{two-point-cft}
\langle T_{ij}(x)T_{kl}(x')\rangle_{CFT} =\frac{C_T}{(x-x')^8}{\cal I}_{ij,kl}(x-x')
\end{equation}
where
\begin{align}
{\cal I}_{ij,kl}(x) =\frac{1}{2}\left(I_{ik}(x)I_{jl}(x)+I_{il}(x)I_{jk}(x)\right)-\frac{1}{4}\eta_{ij}\eta_{kl}   \;\;\;\;  \text{and} \;\;\;\; I_{ij}(x)=\eta_{ij}-2\frac{x_ix_j}{x^2} \,.
\end{align}

According to the AdS/CFT prescription, one should perturb the metric in the bulk as $g_{\mu\nu}\rightarrow g_{\mu\nu}+h_{\mu\nu} $ and solve the corresponding equation of motion for $h_{\mu\nu}$ subject to the boundary condition $h_{\mu\nu}^{(0)}$. Then the quadratic part of the action gives the boundary two-point function of the energy-momentum tensor. Of course this procedure involves divergences and some regularization is needed. The new ingredient is the massive potential (\ref{U1-U4}) which, when multiplied by $\sqrt{-g}$, is divergent in the near boundary limit. However we argue that the overall action doesn't need any mass dependent counterterm. So for our purposes, following \cite{Myers:2010jv}, it is enough to consider $h_{xy}=r^2\phi(r,z)/L^2$ perturbation in the AdS background and find the logarithmic behavior of the action in the momentum space as given in the following. Consider the action in second order of $\phi$,
\begin{equation}\label{action-2point}
I_2=\frac{1}{2\ell_p^3}\int d^5x \left(K_r(\partial_r \phi)^2+K_z(\partial_z\phi)^2+K_m\phi^2+\partial_r\Gamma_r+\partial_z\Gamma_z\right)
\end{equation}
in which
\begin{align}
K_r=&- \frac{r^5\sqrt{f_\infty}}{2L^5}(1-2\lambda f_\infty-3\mu f_\infty^2)  \nonumber\\
K_z=&- \frac{r}{2\sqrt{f_\infty}L}(1-2\lambda f_\infty-3\mu f_\infty^2)  \nonumber\\
K_m=&-\frac{m^2 c_0 r}{4\sqrt{f_\infty}L^3}(3 c_1 r+ 2 c_2 c_0)
\end{align}
where $\Gamma_z$ and $\Gamma_r$ do not contribute to the equation of motion and their contributions to the two-point function are canceled by a generalized Gibbons-Hawking boundary term \cite{Myers:2010jv}\cite{Buchel:2004di}. Take $\phi=e^{i p z}\phi_p(r)$, one finds the equation of motion as,
\begin{equation}
\phi_p''(r)+\frac{5}{r}\phi'_p(r)-\frac{A^2}{r^4}\phi_p(r) -\frac{B}{r^3}\phi_p(r)=0
\end{equation}
where
\begin{align}
B=& \frac{3L^2c_0c_1m^2}{(1-2\lambda f_\infty-3\mu f_\infty^2)} \nonumber\\
A^2=&\frac{L^4 (p^2+\alpha_m^2)}{f_\infty} \nonumber\\
\alpha_m^2=& \frac{2c_0^2c_2m^2}{L^2(1-2\lambda f_\infty-3\mu f_\infty^2)} 
\end{align}
The general solution can be found as
\begin{equation}\label{solution}
\phi_p(r)= b_2\frac{e^{-\frac{A}{r}}}{r^4}  \, _1F_1\left(\frac{B}{2 A}+\frac{5}{2};5;\frac{2 A}{r}\right)+ b_1\frac{e^{-\frac{A}{r}}}{r^4}  U\left(\frac{B}{2 A}+\frac{5}{2},5,\frac{2 A}{r}\right)
\end{equation}
where $U(a,b,x)$ and $_1F_1(a;b;x)$ are the Tricomi and the first kind confluent hypergeometric functions, respectively. Applying the boundary condition of $\phi_p(r_\infty)=1$, we take $b_2=0$ and 
$$ b_1= \frac{1}{6}(A^2-B^2)(9A^2-B^2)\Gamma(\frac{-3}{2}+\frac{B}{2A})$$
Substituting in the action (\ref{action-2point}), and using the equation of motion, we find,
\begin{align}\label{action-two-point}
I_2=&\frac{1}{2\ell_p^3}\int d^5x \partial_r\left(K_r\phi\partial_r \phi  \right)\nonumber\\
=&\frac{b_1^2}{2\ell_p^3}\int d^5x \partial_r\left[K_r U\left(\frac{B}{2 A}+\frac{5}{2},5,\frac{2 A}{r}\right)\partial_r U\left(\frac{B}{2 A}+\frac{5}{2},5,\frac{2 A}{r}\right) \right]
\end{align}
Notice that in the first line, explicit dependence on the mass term disappears. This explains why no new counterterm is needed for the massive potential.
From (\ref{action-two-point}), one can derived the two point function $\langle T_{xy}(x)T_{xy}(x')\rangle $ of the boundary theory. Of course, the latter is not expected to be in conformal form in the presence of the mass term. So we should reach to some deformation of (\ref{two-point-cft}).

{\it i) $c_1=0$: }

For simplicity let us first take $c_1=0$ which corresponds to $B=0$ in (\ref{action-two-point}). The solution to the equation of motion reduces to the modified Bessel function of the second kind with a momentum modification $p^2\rightarrow p^2+\alpha_m^2$ in the solution of \cite{Myers:2010jv}:
\begin{equation}
\phi_p(r)|_{B=0} = \frac{A^2}{2r^2}K_2(\frac{A}{r}) 
\end{equation} 
Now plug in the action and look for $r\rightarrow \infty$ behavior, beside the divergent terms, we find the logarithmic part of the solution (\ref{solution}) as:
\begin{equation}\label{two-point-xspace}
\langle T_{xy}(x)T_{xy}(0)\rangle= \frac{1}{8 f_\infty^{3/2}} \frac{L^3}{\ell_p^3}(1-2\lambda f_\infty-3\mu f_\infty^2)(p^2+\alpha_m^2)^2\left(\log{(p^2+\alpha_m^2)}+N\right)+{\cal O}(\frac{1}{x^2}) 
\end{equation}
where $N$ is a constant number. On the other hand, the Fourier transform of the two-point function  (\ref{two-point-cft}), based on a CFT, read as \cite{Gubser:1997se}:
\begin{align}\label{2point-cft}
\langle T_{xy}T_{xy}\rangle_{(CFT)}(p)=  \frac{C_T}{640}p^4\int d^4x \frac{e^{ip\cdot x}}{x^4} =\frac{\pi^2 C_T}{640}p^4 \log p^2 + \cdots 
\end{align}
where $\cdots$ stands for analytic terms in $p$. Comparison with 
(\ref{two-point-xspace}) shows the replacement of $p^2\rightarrow p^2+\alpha_m^2$ which indicates the deviation from conformal symmetry by the dimensionful parameter $m^2$. In terms of the inverse Fourier transform of (\ref{two-point-xspace}), considering only the logarithmic term, we find\footnote{Here we assumed $\alpha_m^2>0$. For $\alpha_m^2<0$, $\sin(\alpha_m x)$ instead of $\exp(-\alpha_m x)$ appears.} 
\begin{equation}\label{2point-pspace}
\langle T_{xy}T_{xy}\rangle(p)= \frac{1}{8 f_\infty^{3/2}} \frac{L^3}{\ell_p^3}(1-2\lambda f_\infty-3\mu f_\infty^2)p^4\int d^4x \frac{e^{-\alpha_m x}e^{ip\cdot x}}{x^4} 
\end{equation}
where the $p^4$ factor stands for the tensorial part of the correlation function. 
One can read $C_T$ in (\ref{2point-cft}) from (\ref{2point-pspace}) which gives the central charge of the CFT,
\begin{equation}
c=\frac{\pi^4}{40}C_T= \frac{\pi^2}{f_\infty^{3/2}} \frac{L^3}{\ell_p^3}(1-2\lambda f_\infty-3\mu f_\infty^2).
\end{equation}

The exponential factor $e^{-\alpha_m x} $ shows the deviation from conformal invariance by a massive relevant operator with mass $\alpha_m$. 

{\it ii) $c_1\neq 0$: }

In this case, similar calculation gives the logarithmic part of the energy-momentum two point function,
\begin{equation}
\langle T_{xy}(x)T_{xy}(0)\rangle= \frac{1}{72 L^5}\sqrt{f_\infty} (1 - 2 f \lambda - 3 f^2 \mu) 
(9 A^2 - B^2) ( A^2 - B^2) \log{(p^2+\alpha_m^2)}
\end{equation}
In contrast to $B=0$, it is more difficult to find the inverse Fourier transform and extract the correct tensorial behavior. However, the shift in $p^2$ indicates a factor of $e^{-\tilde{\alpha}_m x}$ which in turn predicts presence of a massive operator on the boundary.

\subsection{The $a$-theorem}
In this subsection we are looking for a possible $a$-theorem in the context of the massive gravity. For a moment consider $m=0$ and start from a conformal field theory where in 4-dimensions, it includes two central charges, $c$ and $a$ which can be derived by putting the conformal field theory on a curved background. Then the trace read as, 
\begin{align}\label{trace-anomaly}
&<T_{i}^{i}>=\frac{c}{16\pi^2}I_4- \frac{a}{16\pi^2}E_4\\
&I_4=R_{ijkl }R^{ijkl}-2R_{ij}R^{ij}+\frac{1}{3}R^2\\
&E_4=R_{ijkl }R^{ijkl}-4R_{ij}R^{ij}+R^2
\end{align}  
where $E_4$ and $I_4$ are respectively the Euler density and the Weyl tensor squared. In the standard method for deriving central charges, one considers the Fefferman-Graham expansion of the metric near the boundary as \cite{Henningson:1998gx}
\begin{align}
ds^2=\frac{\tilde{L}^2}{4\rho^2}d\rho^2+\frac{g_{ab}}{\rho}dx^a dx^b
\end{align}
where the boundary is at $\rho=0$ and 
\begin{align}
g_{ab}=g_{(0)ab}+\rho g_{(1)ab} +\rho^2 g_{(2)ab} + \cdots
\end{align}
in which next to the leading terms are determined by the EoM's. However, starting with a general background in a higher curvature theory is very difficult. Instead there is a nice trick by \cite{Sen:2012fc} as taking the boundary metric to be,
\begin{equation}\label{ads2s2}
ds^2=u(1+\alpha \rho) (-r^2 dt^2+ \frac{dr^2}{r^2}) +v(1+\beta \rho)(d\theta^2+\sin^2\theta d\phi^2)
\end{equation} 
This is indeed the $AdS_2\times S^2$ background with $\alpha$ and $\beta$ as perturbations which can be found from the EoM. It is known that the trace anomaly is found by the logarithmic part of the action as follows
\begin{equation}
I_{Ln}=-\frac{1}{2}\int \sqrt{g_0} <T_{i}^{i}>
\end{equation}
where $Ln$ subscript means the logarithmic divergent part. Now simply extremize the action with respect to  $\alpha$ and $\beta$ and plug them back to the action, 
one can easily read the central charges as coefficients of $I_4$ and $E_4$ where they appear as
\begin{equation}
I_4=\frac{4}{3}\left(\frac{1}{u^2}+\frac{1}{v^2}-\frac{2}{uv}\right)\;, \hspace{1cm} E_4=-\frac{8}{uv} \,\,.
\end{equation}
So final form of the trace anomaly would be
\begin{equation}\label{TT}
<T_{i}^{i}>=\frac{c}{12\pi^2}\left(\frac{1}{u^2}+\frac{1}{v^2}-\frac{2}{uv}\right)+\frac{a}{2\pi^2}\frac{1}{uv}
\end{equation}
By this procedure, central charges can be found  as \cite{Myers:2010jv},
\begin{align}\label{central-1}
	c=&  \frac{\pi^2}{f_\infty^{3/2}} \frac{L^3}{\ell_p^3} (1 - 2 \lambda f_\infty - 3 \mu f_\infty^2) \nonumber\\
	a=& \frac{\pi^2}{f_\infty^{3/2}} \frac{L^3}{\ell_p^3} (1 - 6 \lambda f_\infty + 9 \mu f_\infty^2)
\end{align}

Now consider the massive gravity. At the first glance introducing a mass scale seems to ruin the scale invariance of the theory and one may expect a non-zero energy-momentum trace because of explicit scale symmetry breaking. This is in addition to the conformal trace anomaly and some deviation from the functional form of (\ref{TT}) is expected. Now the question is how the flow behavior out of the conformal fixed point. This can be studied by the so-called $c$ or $a$-theorem of \cite{Myers:2010tj} and we are going to consider it in the context of massive gravity. 

In any $a$-theorem, the task is to find a monotonically decreasing function of the scale which matches with the central charge at the fixed point. In holography, most of $a$-theorems are based on the null energy condition in the bulk theory \cite{Myers:2010xs,Liu:2010xc,Liu:2011iia}. Let us consider it for the massive gravity. We start with the following metric,
\begin{equation}
ds^2=e^{2A(r)}\left(-dt^2+d\vec{x}^2_{d-1}\right)+dr^2
\end{equation}
In the large $r$ limit, we assume $A(r)=r/\tilde{L}$ and the metric becomes asymptotically $AdS$. Suppose a theory consisted of the action (\ref{Action}) as the gravity part with some matter source not included there, then the generalized Einstein equation has the following form,
\begin{equation}
G_{\mu\nu}-\frac{6}{L^2} g_{\mu\nu}+ H_{\mu\nu}+Z_{\mu\nu}+M_{\mu\nu}= T_{\mu\nu}
\end{equation}
where $G$ is the Einstein tensor and $H$, $Z$ and $M$ are respectively variations of Gauss-Bonnet, quasi-topological and massive terms. 

Now let us turn to the reference metric which indeed can be considered as a sort of coupling in the action. We take $f_{\mu\nu}$ to be fixed and the same as the beginning of this section,  
\begin{equation}
f_{ab}dx^a dx^b=\frac{c_0^2}{L^2}\left(dx_1^2+dx_2^2+dx_3^2\right) \,.
\end{equation}
Now we introduce the $a$-function through,
\begin{align}\label{nec-1}
a'(r)=& -\frac{3\pi^2}{\ell_p^3 A'(r)^4}A''(r)\left(1-2\lambda L^2 A'(r)^2-3 \mu L^4 A'(r)^4\right)  \nonumber\\
=& -\frac{\pi^2}{\ell_p^3 A'(r)^4}\left(T^t_t-T^r_r\right)\geq 0.
\end{align}
The second equality comes from the equation of motion and the inequality indicates the null energy condition for the matter field energy-momentum tensor. A simple integration of the above function gives
\begin{align}
a(r)=& \frac{\pi^2}{\ell_p^3 A'(r)^3}\left(1-6\lambda L^2 A'(r)^2+9 \mu L^4 A'(r)^4\right) 
\end{align}
In the UV limit, we have the $AdS$ background where $A(r) \sim r/\tilde{L}$, then the $a$ central charge previously introduced in (\ref{central-1}) will be recovered which is exactly the Euler density coefficient in the trace anomaly (\ref{trace-anomaly}). Notice that this analysis is in the presence of the mass term in the action and gives the same functional form as in the literature for massless case \cite{Myers:2010tj}. This follows from the fact that the null energy condition used in (\ref{nec-1}) was based on $\xi_{(1)}^{\mu}\xi_{(1)}^{\nu}T_{\mu\nu}\geq 0$ with $\xi_{(1)}=(e^{-A(r)},0,0,0,1)$ which has no component in non-vanishing directions of the reference metric. 


\section{Quasi-Topological Black Brane in Gauss-Bonnet Massive Gravity}
\label{sec-blackbrane}
In this section, we study a black brane solution to the action (\ref{Action}). We derive its metric, temperature and the entropy, then compute the rate of entropy production due to a strain which in the massless limit corresponds to viscosity of the hydrodynamic limit of the dual theory on the boundary.

\subsection{The black brane solution}
We consider the following ansatz for the metric of five-dimensional planar AdS black brane,
\begin{equation}\label{metric-fr}
ds^{2} =-\frac{r^2N(r)^2}{L^2}f(r)dt^{2} +\frac{L^2dr^{2}}{r^2f(r)} +r^2h_{ij}dx^idx^j,
\end{equation}
A generalized version of the reference metric $ f_{\mu \nu} $ was proposed in {\cite{deRham:2010kj}} with the  form
$ f_{\mu \nu} = diag(0,0,c_0^2h_{ij})$, where $ h_{ij}=\frac{1}{L^2}\delta_{ij} $.

The values of $ \mathcal{U}_i $ in (\ref{7}) are calculated as below,
\begin{align}\label{9} 
  & \mathcal{U}_1=\frac{3c_{0}}{r}, \,\,\,  \,\,\, \mathcal{U}_2=\frac{6c_0^2}{r^2},\,\,\,\,\mathcal{U}_3=\frac{6c_0^3}{r^3},\,\,\,\,\mathcal{U}_4=0 
   \end{align} 
Inserting this  ansatz into the action (\ref{Action}) yields,
\begin{equation}\label{EOMI}
 I =\frac{1}{2\ell_p^3}\int{d^5x\frac{3N(r)}{L^5}\frac{d}{dr}\Bigg[r^4\Bigg(1-f(r)+\lambda f(r)^2+\mu f(r)^3+\frac{\Upsilon(r)}{r^4}\Bigg)\Bigg]}
\end{equation}
in which 
\begin{align}
\Upsilon(r)=r_0^4\Bigg(&\frac{1}{3}m_1\frac{r^3}{r_0^3}+m_2\frac{r^2}{r_0^2}+2m_3\frac{r}{r_0}\Bigg)  \\
m_1=\frac{m^2L^2c_0c_1}{r_0},\;\;\;m_2&=\frac{m^2L^2c_0^2c_2}{r_0^2}, \;\;\; m_3=\frac{m^2L^2c_0^3c_3}{r_0^3} 
\end{align}
where $m_i$'s are dimensionless mass parameters.\\
By variation of $ N(r) $ we have {\cite{Myers:2010ru}},
\begin{equation}
\frac{d}{dr}\Bigg[r^4\Bigg(1-f(r)+\lambda f(r)^2+\mu f(r)^3+\frac{\Upsilon}{r^4}\Bigg)\Bigg]=0
\end{equation}
$ f $ is given by solution of the following equation,
\begin{equation}\label{blackening-equ}
r^4\Bigg(1-f(r)+\lambda f(r)^2+\mu f(r)^3+\frac{\Upsilon}{r^4}\Bigg)=b^4
\end{equation}
in which $b$ is a constant of motion and can be determined as a function of the radius of horizon at which $f(r_0)=0$:
\begin{equation}
b^4=r_0^4+\Upsilon_0
\end{equation}
where $\Upsilon_0=\Upsilon(r_0)$, then
\begin{equation}\label{f-equation}
1-f(r)+\lambda f(r)^2+\mu f(r)^3=\frac{r_0^4}{r^4}+\frac{\Upsilon_0-\Upsilon(r)}{r^4}
\end{equation}

That's easy to show $ N(r) $ is constant by variation of $ f(r) $ from (\ref{EOMI}). The speed of light in the boundary CFT is simply $c=1$, thus we have $\lim_{r \to \infty}{N^2 f(r)}=1$ so we take $N=1/\sqrt{f_\infty}$. \\

The temperature and the Hawking-Bekenstein entropy density can be found as \cite{Bekenstein},
\begin{align}\label{temperature}
	T&=\frac{1}{2\pi}[\frac{1}{\sqrt{g_{rr}} } \frac{d}{dr} \sqrt{-g_{tt}}]|_{r=r_0 } =\frac{r_0^2}{4\pi L^{2}\sqrt{f_\infty}}\frac{df}{dr}|_{r=r_0} =\frac{r_0}{\pi L^2\sqrt{f_\infty}}(1+M_1) \\
	s&=\frac{1}{2\ell_p^3}\frac{4\pi}{V}\int d^{3}x \sqrt{-g}=2\pi\left(\frac{r_0}{\ell_p L}\right)^{3}.
\end{align}
where in the first line we used (\ref{f-equation}) to find $\frac{df}{dr}|_{r=r_0}$ with
\begin{equation}
	M_1=\frac{1}{4}\left(m_1+2m_2+2m_3\right)
\end{equation}

\subsection{The rate of entropy production}
As explained in the introduction, the concept of viscosity as a hydrodynamic transport coefficient is ill-defined when translational invariance is violated. Authors of \cite{Hartnoll:2016tri} suggested that the right quantity to be the rate of entropy production per Planckian time  due to a strain (e.g. $\delta g_{xy}$) and can be derived from the Kubo formula as,
\begin{align}\label{Kubo}
\eta=\lim_{\omega\rightarrow 0}\frac{1}{\omega}\mathcal{G}^R_{T_{xy}T_{xy}}
\end{align}     
where $\mathcal{G}^R$ is the retarded Green's function for the stress tensor.
Nonetheless, finding this quantity which in the following for simplicity we call it the shear viscosity, is a challenge in any massive gravity. While in a massless theory, the translational invariance implies a trivial flow for the metric perturbation and leads to a simplification by which the calculation can be done completely in terms of the horizon information \cite{Iqbal:2008by}. In contrast, viscosity was found in massive gravity in the Einstein \cite{Hartnoll:2016tri} (see also \cite{Pan:2016ztm}) and Gauss-Bonnet theories \cite{Sadeghi:2015vaa}. The calculation in \cite{Hartnoll:2016tri} is based on the analysis in \cite{Lucas:2015vna} where the retarded Green's function was found for a massive scalar field in the bulk. The result can be written as 
\begin{equation}
\frac{\eta}{s} = \frac{1}{4\pi} \phi_0^2
\end{equation}
where $\phi_0$ is the zero frequency of the metric perturbation $\delta g_x^y$ at the horizon. For a massless field the translational invariance implies a trivial flow from horizon to boundary. The massless scalar field is therefore a constant everywhere and equals to its boundary value $\phi_0=1$. So the KSS bound is saturated in this case as $\eta/s=1/(4\pi)$. However, in the massive case 
the translational symmetry no longer exists and one expects a nontrivial $\phi_0$. In \cite{Hartnoll:2016tri}, $\phi_0$ was found both perturbatively and numerically for massive gravity in 4 dimensions. It was shown that for the range of parameters where the model is stable, $\phi_0^2$ factor is smaller than one which indicates the violation of the KSS bound. 

In our previous work in \cite{Sadeghi:2015vaa}, we calculated the viscosity in the massive Gauss-Bonnet gravity by the direct application of Green-Kubo formula, of course in the special case of $m_1=m_2=0$ while $m_3\neq 0$. We are now going to generalize the formalism to include more general massive higher curvature theories with two derivatives equation of motions. We take the advantage of pole method of \cite{Paulos:2009yk} which can be generalized for a massive field as we discuss shortly.

Study of the hydrodynamics for a massive field was considered in \cite{Lucas:2015vna}. The analysis is based on the following second order action which can be derived from the original higher curvature action by considering a suitable metric perturbation as $\delta g_x^y=\phi(r)e^{i\omega t}$,
\begin{equation}\label{eff-action}
S=\int d^5 x \frac{\sqrt{-g}}{q(r)}(g^{rr} \phi'^2+E_2\phi^2)
\end{equation}
where $'$ indicates the derivative with respect to $r$ and $E_2$ is the summation of a term in order of $\omega^2$ and a mass term. Here we provide $q(r)$ as an effective $r$-dependent coupling which encodes the higher curvature nature of the theory. This is in the spirit of \cite{Iqbal:2008by}. Following the procedure of \cite{Lucas:2015vna}, the final result for the viscosity is as follows,
\begin{equation}\label{our-recipe}
\frac{\eta}{s} = \frac{1}{4\pi q(r_0)} \phi_0(r_0)^2
\end{equation}
where $\phi_0$ denotes the zero frequency mode and is computed at the horizon. The new ingredient is the effective coupling $q(r_0)$. For a Gauss-Bonnet gravity it is well-known to be $1/q(r_0)=(1-4\lambda)$ (see also \cite{Iqbal:2008by}). 

To compute this effective coupling in a higher derivative gravity, one can use the prescription of pole method in \cite{Paulos:2009yk} which was originally devoted to the massless theories, since it relies on finding the viscosity by a trivial flow and only depends on the horizon information. In our prescription, the pole method is used to find the effective coupling and it works at least for higher curvature theories with second order differential equation of motion, e.g. Gauss-Bonnet, cubic or higher quasi-topological theories of gravity \cite{Dehghani:2013ldu}. This statement comes from the fact that effective coupling in (\ref{eff-action}) represents the higher curvature nature of gravity and does not explicitly depends on the mass term. It therefore can be found by setting $m=0$ and use the pole method for the viscosity. It follows that
\begin{equation}\label{eta0}
\eta_0=-8\pi T \lim\limits_{\omega,\epsilon\rightarrow 0}\frac{Res_{z=0}{\cal L}_0}{\omega^2 \epsilon^2}
\end{equation}
where index $0$ means the zero mass limit and  $z=1-r_0^2/r^2$ is a new coordinate, so that the horizon and boundary are at $z=0$ and $z=1$, respectively. Comparing (\ref{eta0}) and (\ref{our-recipe}) one finds that the effective coupling can be found from the pole method as given in (\ref{effective-coupling}).
 
The procedure is as follows. Take the metric perturbation to be $\phi=\epsilon\phi_0(z)e^{i\omega t}$. The general form of the second order action for second derivative theories are
\begin{equation}\label{K-action}
S=\int d^5x (K_1\phi'^2+K_2\phi^2)
\end{equation}
It is important to note that the effective coupling is encoded in $K_1$ term which its dependence on mass parameter is implicit. Whereas the massive term is explicit in $K_2$ term. So if we follow the prescription of \cite{Paulos:2009yk} we find,
\begin{align}\label{effective-coupling}
\frac{1}{q(0)}&=-16\pi T \left(\frac{\ell_p L}{r_0}\right)^{3} \lim\limits_{\omega,\epsilon\rightarrow 0}\frac{Res_{z=0}{\cal L}_0}{\omega^2 \epsilon^2} 
\end{align}
where ${\cal L}_0$ indicates the total Lagrangian excluding the $\sqrt{-g}U$ term,
\begin{equation}
{\cal L}_0=\sqrt{-g} \Bigg(R+\frac{12}{L^2}+\frac{\lambda  L^2}{2}\mathcal{L}_{GB}+\frac{7}{8}L^4\mu \mathcal{L}_{3}\Bigg)
\end{equation}

Let us apply this prescription to our theory. Firstly, the unperturbed metric read as 
\begin{equation}
	ds^{2} =\frac{r_0^2}{L^2(1-z)}\left(-\frac{f(z)}{f_\infty}dt^{2}+dx_1^2+dx_2^2+dx_3^2\right) +\frac{L^2}{4f(z)}\frac{dz^{2}}{(1-z)^2}
\end{equation}
then we can perturb the metric by the following shifting:
$$ dx_1 \rightarrow dx_1 + \epsilon \phi dx_2$$
The corresponding perturbed metric follows,
\begin{equation}
	ds^{2} =\frac{r_0^2}{L^2(1-z)}\left(-\frac{f(z)}{f_\infty}dt^{2}+dx_1^2+(1+\epsilon^2\phi^2)dx_2^2+2\epsilon\phi dx_1 dx_2+dx_3^2\right) +\frac{L^2}{4f(z)}\frac{dz^{2}}{(1-z)^2}	 
\end{equation}
then,
$$
\mathcal{K}=\frac{c_0\sqrt{1-z}}{r_0}\left(
\begin{array}{ccccc}
0 & 0 & 0 & 0 & 0 \\
0 & 0 & 0 & 0 & 0 \\
0 & 0 &  \left(1+\frac{3}{8} \epsilon^2 \phi^2\right) & -\frac{1}{2} \epsilon  \phi & 0 \\
0 & 0 & -\frac{1}{2}  \epsilon  \phi &  \left(1-\frac{1}{8} \epsilon^2 \phi^2\right) & 0 \\
0 & 0 & 0 & 0 & 1 \\
\end{array}
\right)
$$
and the massive term in the Lagrangian up to second order is,
\begin{align}\label{U-z}
\sqrt{-g} \mathcal{U}=& \frac{3H}{2}\left(m_1+2m_2\sqrt{1-z}+2m_3(1-z)\right)  +\frac{H}{8}\left(m_1+2m_2\sqrt{1-z}\right)\epsilon^2\phi^2+O(\epsilon )^3
\end{align}
where 
\begin{align}
	H=\frac{r_0^4}{L^5\sqrt{f_\infty}(1-z)^{5/2}} \;.
\end{align}
Taking $\phi=\phi_0(z)\exp^{-i\omega t}$, the shear viscosity can be found from (\ref{our-recipe}) and (\ref{effective-coupling}) as follows
\begin{align}\label{our}
\eta&=-8\pi T \lim\limits_{\omega,\epsilon\rightarrow 0}\frac{Res_{z=0}{\cal L}_0}{\omega^2 \epsilon^2}\phi_0(0)^2 \\
&=-8\pi T \lim\limits_{\omega,\epsilon\rightarrow 0}\left[\frac{Res_{z=0}\sqrt{-g} \Bigg(R+\frac{12}{L^2}+\frac{\lambda  L^2}{2}\mathcal{L}_{GB}+\frac{7}{8}L^4\mu \mathcal{L}_{3}\Bigg)}{\omega^2 \epsilon^2} \right]\phi_0(0)^2
\end{align}
It can be found as
\begin{align}\label{eta-f-relation}
\eta&= \frac{r_0^3}{2 \ell_p^3L^3}\left[1-2\lambda f_0'-9\mu(f_0^{'2}+2f_0^{''2}+2f'_0(f^{'''}_0-3f^{''}_0))\right]\phi_0(0)^2
\end{align}
where $f^{(n)}_0=d^n f(z)/dz^n |_{z=0}$ are derivatives of $f$ at horizon. The above equation is the same as \cite{Myers:2010jv} up to $\phi_0(0)^2$ factor. This is natural, since as stated above it includes the effective coupling and does not explicitly depend on mass parameter which is implicit in $f$'s derivatives. In \cite{Myers:2010jv} the $\phi_0(0)^2$ factor is one due to translational invariance in the massless theory.

Let us forget the $\phi_0(0)^2$ factor for a while and derive the $f$ derivatives in (\ref{eta-f-relation}) from (\ref{f-equation}) after rewritten in $z$-coordinate,
\begin{equation}
f(z)-\lambda f^2(z) -\mu f^3(z) = z(2-z)\left(1+\frac{\Upsilon_0}{r_0^4}\right) -\frac{\Upsilon_0}{r_0^4} + {\cal Z}(z) 
\end{equation}where
\begin{align} 
{\cal Z}(z)\equiv& \frac{(1-z)^2}{r_0^4}\Upsilon(z) \nonumber\\
=& \frac{1}{3}m_1\sqrt{1-z} +m_2(1-z)+2m_3 (1-z)^{3/2}
\end{align}
then
\begin{align}
f'_0&=2\left(1+\frac{\Upsilon_0}{r_0^4}\right)+{\cal Z}'_0 = 2(1 + M_1)  \nonumber\\
f_0^{''}&=2\lambda f^{'2}_0 -2\left(1+\frac{\Upsilon_0}{r_0^4}\right)+{\cal Z}^{''}_0 = 2\lambda f^{'2}_0 - 2 + M_2  \nonumber\\
f_0^{'''}&=6\lambda f'_0 f^{''}_0 + 6 \mu f_0^{'3}+{\cal Z}^{'''}_0 = 6\lambda f'_0 f^{''}_0 + 6 \mu f_0^{'3} + M_3  \nonumber\\
\end{align}
in which
\begin{equation}
M_1=\frac{1}{4}\left(m_1+2m_2+2m_3\right), \;\;\; M_2=\frac{1}{4}\left(3m_1+8m_2+10m_3\right), \;\;\;
M_3=\frac{1}{8}\left(-m_1+6m_3\right)   
\end{equation}
then 
\begin{align}
\frac{4\pi\eta}{s}&=\Big[1 - 4(1 + M_1) \lambda 
-36 \mu \Big(9 + 8M_1+M_1^2-5M_2-3M_1M_2+\frac{M_2^2}{2}+M_3+M_1M_3  \nonumber\\
&-4 (1+M_1)^2 (6 M_1-5 M_2+16)\lambda+128 (1+M_1)^4\lambda^2+48 (1+M_1)^4 \mu\Big)\Big]\phi_0(0)^2 \nonumber\\   
\label{etaovers}
&=\Big[ 1 - 4\tilde{\lambda} 
-36 \mu \Big(9 + 8M_1+M_1^2-5M_2-3M_1M_2+\frac{M_2^2}{2}+M_3+M_1M_3  \nonumber\\
&-4 (1+M_1) (6 M_1-5 M_2+16)\tilde{\lambda}+128 (1+M_1)^2\tilde{\lambda}^2+48 (1+M_1)^4 \mu\Big)\Big]\phi_0(0)^2
\end{align}
with $\tilde{\lambda}=(1+M_1)\lambda$. Taking massless limit by $M_i=0$ for which we know $\phi_0(0)^2=1$ and get,
\begin{equation}\label{Mayers-result}
\frac{4\pi\eta}{s}=1-4 \lambda -36\mu (9-64 \lambda + 128 \lambda^2 +48 \mu ^2)
\end{equation}
which is the same as result of \cite{Myers:2010jv}. If otherwise we put $\mu=0$ in (\ref{etaovers}),
\begin{align} 
\frac{4\pi\eta}{s}&=(1 - 4\tilde{\lambda})\phi_0(0)^2=(1 - 4(1 + M_1) \lambda)\phi_0(0)^2 \nonumber\\
&=( 1-  \frac{4\pi L^2 \sqrt{f_\infty} T}{r_0} \lambda)\phi_0(0)^2
\end{align}
This matches with \cite{Sadeghi:2015vaa} up to $\phi_0(0)^2$ factor. We will comment on this discrepancy soon.\\
Checking this result for Einstein massive gravity  we set $\lambda=\mu=0$ in (\ref{etaovers}) and find $\eta/s=\phi_0(0)^2/(4\pi)$. The result is the same as \cite{Hartnoll:2016tri}.\footnote{In \cite{Pan:2016ztm}, applying Petrov-like boundary condition leads to the KSS saturated bound $\eta/s=1/4\pi$. Here we are using the Dirichlet boundary condition and regularity on horizon which corresponds to \cite{Hartnoll:2016tri}.}

{\bf Finding $\phi_0(0)^2$:}\\
It is now time to find $\phi_0$ which is solution of the following second order equation derived from action (\ref{K-action}):
\begin{equation}\label{K-equation}
(K_1\phi_0')'-K_2\phi_0=0
\end{equation}
where $'$ denotes derivative with respect to $z$ and 
\begin{align}\label{K1K2}
K_1(z)&= \frac{8 f(z)}{1-z}
 \Big[-1 + 3 \mu f(z)^2 + 
9 (z-1)^2 \mu f'(z)^2 + 
2 f(z) \Big(\lambda - 3 (z-1) \mu f'(z)\Big) \nonumber\\
&+18(z-1)^4 \mu f''(z)^2 + 
2 (z-1)f'(z) \Big(\lambda - 27 (z-1)^2 \mu f''(z) - 
9 (z-1)^3 \mu f^{(3)}(z)\Big)\Big]  \nonumber\\
K_2(z)&=\frac{m_1\sqrt{1-z}+2m_2(1-z)}{(1-z)^3}
\end{align}
where we put $\omega=0$ in $K_2$. $\phi_0$ is subjected to two boundary conditions which are the regularity at the horizon and $\phi_0(1)=1$ on the boundary.\\
Let us firstly consider a special case where $m_1=m_2=0$ while $m_3\neq 0$. This gives $K_2=0$ and the only regular solution is the constant one $\phi_0(0)=1$. This matches exactly with our previous work \cite{Sadeghi:2015vaa} for massive Gauss-Bonnet gravity in which we considered $m_1=m_2=0$.\\
Nonetheless, for a general case it would be very hard, if not impossible to find an exact solution or any shortcut to find the horizon value of $\phi_0(0)$. So we perform a numeric solution. It will be helpful to look at near the horizon behavior. Regarding the regularity of $\phi_0(z)$, we consider the following Taylor's expansions around $z=0$. 
\begin{align}\label{expansion}
\phi_0(z)&= a_0+a_1 z+\frac{1}{2} a_2 z^2 +\cdots, \nonumber\\
K_1(z)&= b_1 z+\frac{1}{2} b_2 z^2 +\cdots,\nonumber\\
K_2(z)&= d_0+d_1 z+\frac{1}{2} d_2 z^2 +\cdots.
\end{align}
where $a_0=\phi_0(0)$ and $b_i$'s and $d_i$'s can be read from (\ref{K1K2}).
Inserting (\ref{expansion}) expansions into equation of motion (\ref{K-equation}), one finds
\begin{align}
a_1&=\frac{d_0}{b_1}a_0  \nonumber\\
a_2&=\frac{b_1d_1+d_0^2-b_2d_0}{2b_1^2}a_0 
\end{align}
These two coefficients help us in applying the boundary condition at the horizon for our numerical analysis. We also recall that the couplings are constrained by unitarity, causality, and positivity of energy fluxes in the dual conformal field theory. It turns out that it must be \cite{Myers:2010jv},
\begin{align}
-7/36\lesssim&\,\,\lambda_{gb}\,\,\lesssim 9/100\nonumber\\
&\vert{\mu}\vert<0.001
\end{align} 
\begin{figure}[h]\label{figure1}
	\centering
	\includegraphics[height=50mm]{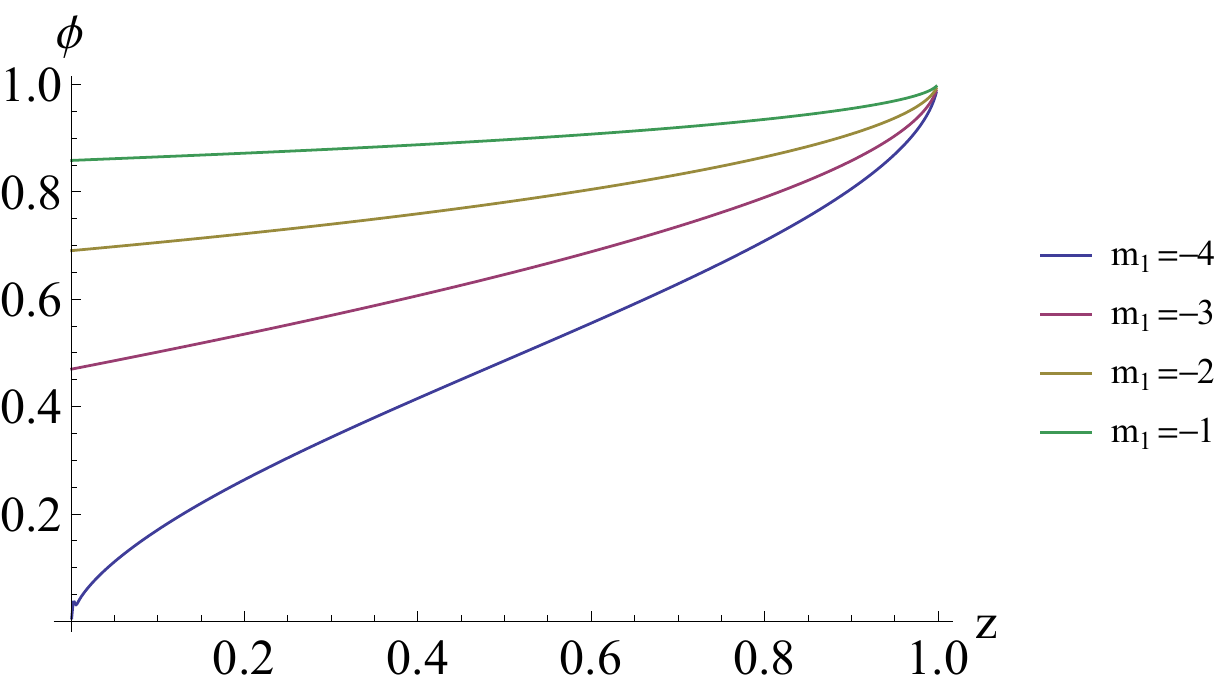} \\
	\caption{$\phi_0(z)$ versus $z$ between the horizon and boundary, $0\leq z \leq 1$.  We set $\mu=0.001$ and $m_2=m_3=0$. $m_1$ varies from -4 to -1 from bottom to top.}	
\end{figure}
Figure 1 displays the profile of $\phi_0(z)$ from the horizon to boundary for different values of $m_1$ where by (\ref{temperature}) and setting $m_2=m_3=0$, it indeed shows different temperatures with $m_1=-4$ corresponding to zero temperature.

Figure 2 shows the $\phi_0(0)^2$ factor and $4\pi\eta/s$ as functions of dimensionless mass parameter $m_1$ with different Gauss-Bonnet coupling $\lambda$. It indicates that for negative $m_1$ which is in the physical range (see the next section), the $\phi_0(0)^2$ factor is less than one, while for positive $m_1$ is greater than one irrespective of the sign of $\lambda$. This is in agreement with results and general arguments given by \cite{Hartnoll:2016tri}.
\begin{figure}[h]\label{figure2}
	\centering
	\includegraphics[height=50mm]{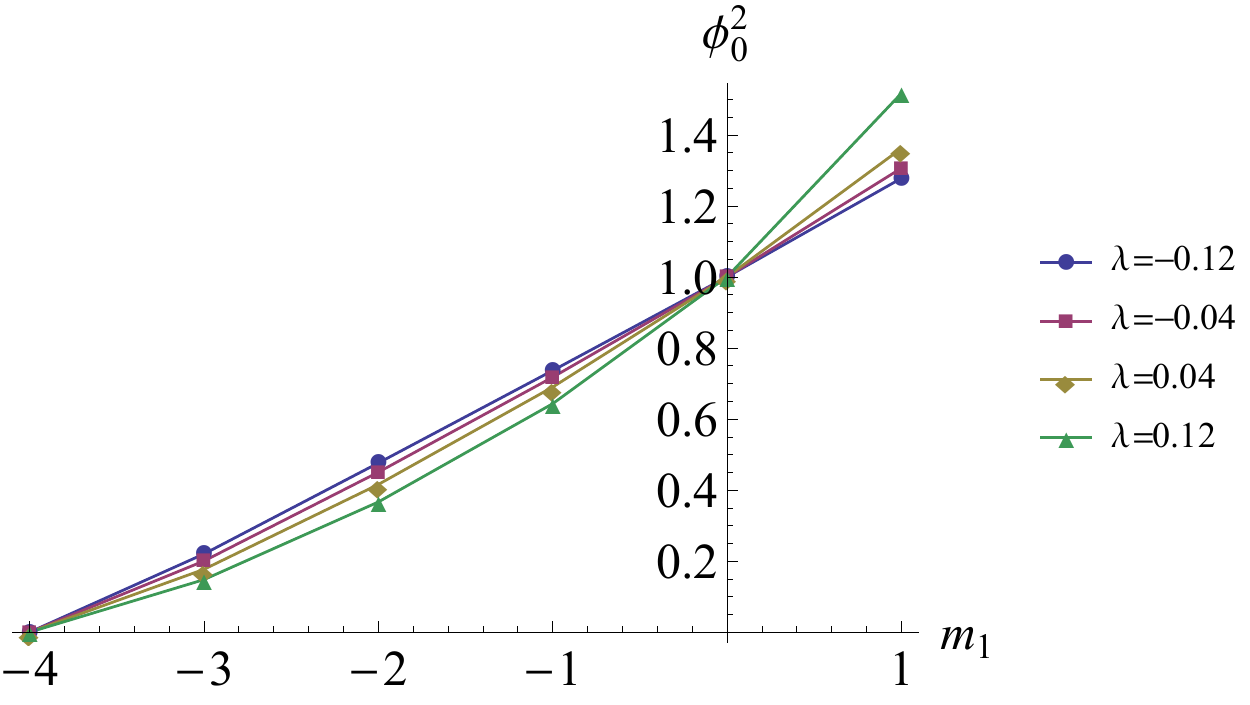}\\
	\includegraphics[height=50mm]{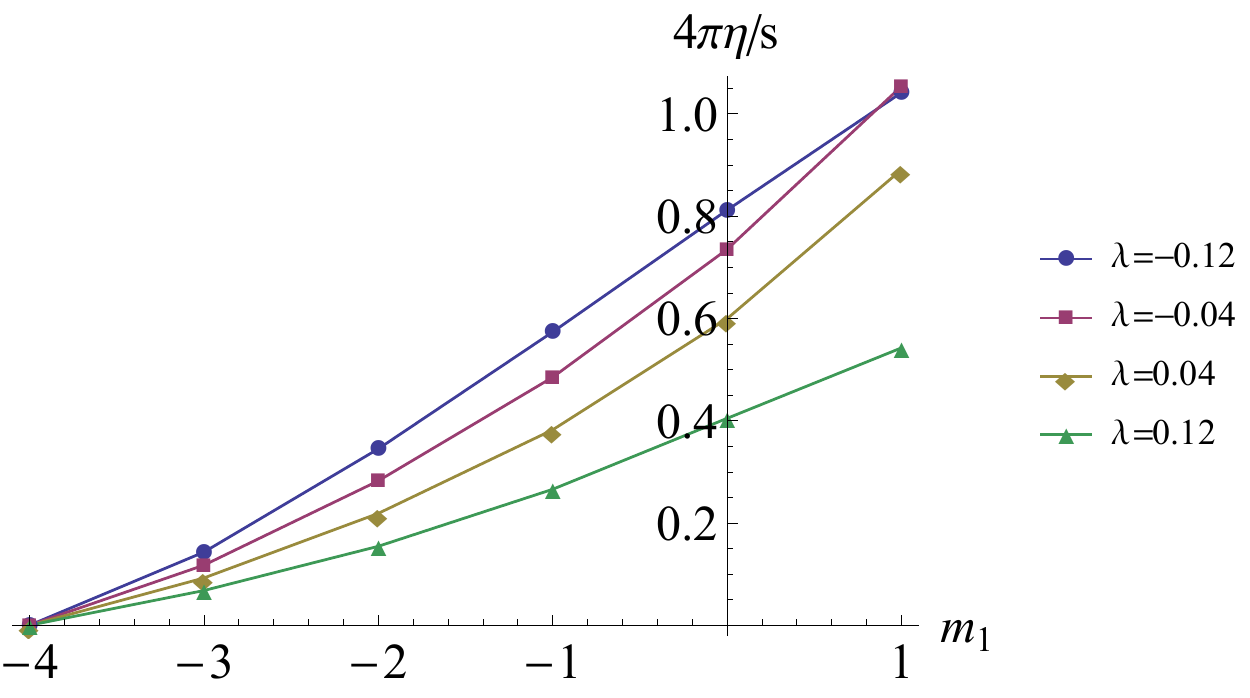} \\
	\caption{$\phi_0(0)^2$ factor (top) and $4\pi\eta/s$ (bottom) as functions of dimensionless mass parameter $m_1$ with different Gauss-Bonnet coupling $\lambda$. We set $\mu=0.001$ and $m_2=m_3=0$.}	
\end{figure}

\section{Physical constraints}

Here we check some physical conditions on our parameters. The first one is unitarity which means that the norm of the energy momentum two-point function (\ref{2point-pspace}) is positive. Therefore, the $c$-central charge should be positive, then
\begin{align}\label{unitarity}
1-2\lambda f_\infty-3\mu f_\infty^2>0
\end{align}
We have from (\ref{f-equation})
\begin{align}
\tilde{\Gamma}(f)\equiv &1-f(r)+\lambda f(r)^2+\mu f(r)^3-\frac{r_0^4}{r^4}-\frac{\Upsilon_0-\Upsilon(r)}{r^4}\\
-\tilde{\Gamma'}(f_\infty)=&1-2\lambda f_\infty-3\mu f_\infty^2>0
\end{align}
This is exactly the ghost free condition for the graviton propagating on an AdS background \cite{Myers:2010ru}. 

The next constraint comes from the causality in the CFT. This is nontrivial since the 4-dimensional Lorentz symmetry is broken by the black hole background as well as by the reference metric. To investigate this, we consider the front velocity of signals in the dual CFT  determined by $v^{front}=\mathop{\lim }\limits_{|q| \to \infty} \frac{Re(\omega)}{q}$ which corresponds to the phase velocity of the short wavelength modes. So the necessary condition for causal behavior is $v^{front}\leqslant1$. Define a new coordinate $\rho=\frac{{r_{0}}}{r}$ the metric  (\ref{metric-fr}) becomes as follows,
\begin{equation}\label{metric-rho}
ds^{2} =\frac{{r_0}^2}{L^2\rho^2}\Big(-\frac{f(\rho)}{f_\infty}dt^{2}+dx_1^2+dx_2^2+dx_3^2\Big) +\frac{L^2d\rho^{2}}{\rho^2f(\rho)} 
\end{equation}
One can find linearized equations of motion  by perturbing the metric with $h_{x^1x^2}=\frac{r_0^2}{L^2\rho}e^{-i\omega t+iqx_3}\phi(\rho), $  as follows,
\begin{equation}
\partial_\rho\Big(C^{(2)}(\rho,q^2)\partial_\rho \phi(\rho)\Big)+C^{(0)}(\rho,q^2,\omega^2)\phi(\rho) +C^{(m)}(\rho,m^2)\phi(\rho)=0
\end{equation}
The radial derivatives can be ignored  since we are in the large momentum and frequency limit. Moreover the $C^{(m)}$ term comes from the mass term which doesn't involve any derivative so independent of $q^2$ and $\omega^2$. We therefore ignore the $C^{(m)}$  as well. In this way, we find linearized equation as in the massless case \cite{Myers:2010jv}\footnote{Notice we differ from \cite{Myers:2010jv} by choosing $\rho=\frac{{r_{0}}}{r}$ instead of $\rho=\frac{{r_{0}}^2}{r^2}$. A simple change of variable transforms our linearized equations to those of \cite{Myers:2010jv}.}. It reduces to the following equation where only terms proportional to $q^2$, $\omega^2$ and their higher degrees are survived in the short wavelength limit,  
\begin{align}\label{casuality-eom}
&0 = \omega^2 \Big( 1 - 2\lambda f(\rho) +\rho \lambda f'(\rho) \Big)-\frac{ f(\rho)}{f_\infty}q^2\Big( 1 - 2\lambda f(\rho) +2\rho \lambda f'(\rho) -\rho^2 \lambda f''(\rho)\Big)\nonumber\\
&-3\mu \omega^2\Big[ f(\rho)\Big( f(\rho) - \rho f'(\rho)  +\frac{3}{2}\rho^3f^{(3)}(\rho)+\frac{3}{4}\rho^4f^{(4)}(\rho)\Big)\nonumber\\
&+\frac{3}{8}\rho^2 f'(\rho) \Big( \rho f''(\rho) +\rho^2  f^{(3)}(\rho) \Big)+\frac{3}{8}\rho^4 f''(\rho)^2\Big]\nonumber\\
&+3\mu\frac{f(\rho)}{f_{\infty}}q^2\Big[ f(\rho)\Big( f(\rho) - 2 \rho f'(\rho) +\rho^2 f''(\rho) -\frac{3}{2}\rho^3f^{(3)}(\rho)-\frac{3}{4}\rho^4f^{(4)}(\rho)\Big)\nonumber\\
&\frac{1}{4}\rho^2 f'(\rho) \Big(4 f'(\rho) -3\rho f''(\rho) -3\rho^2  f^{(3)}(\rho) \Big)\Big]\nonumber\\
&-6\mu \rho^2 \frac{f(\rho)}{f_{\infty}}q^2 f''(\rho)\Big(\omega^2-\frac{f(\rho)}{f_{\infty}}q^2\Big)
\end{align}
Let us proceed step by step from the Einstein gravity where $\lambda$ and $\mu$ vanish. We thus have, 
\begin{equation}
v_f^2\equiv\lim_{q^2\rightarrow\infty}\frac{\omega^2}{q^2}=\frac{f(\rho)}{f_{\infty}},
\end{equation}
where $f_\infty=1$ and 
\begin{equation}
f(\rho)=1 + \frac{1}{3}m_1 \rho + m_2 \rho^2 + 2 m_3 \rho^3 - \rho^4
\end{equation}
For massless case it satisfies the causality condition, while in the massive gravity, the dominant term near the boundary is $m_1$ and should be negative to achieve the causality  $v_f^2 <1$. One may set $m_1=0$, then $m_2$ should be nonpositive and so on for $m_3$.

In the second step, consider the Gauss-Bonnet gravity, {\it i.e.} $\lambda\neq 0$ and $\mu=0$. Then only the first line of (\ref{casuality-eom}) is nonvanishing. We may insert 
\begin{equation}
f(r)=f_\infty + f'(0) \rho + \frac{1}{2} f''(0) \rho^2 + \frac{1}{6} f^{(3)}(0) \rho^3 + 
\frac{1}{24} f^{(4)}(0)\rho^4+\cdots
\end{equation}
where $f$ derivatives can be found by (\ref{blackening-equ}) rewritten in $\rho=r_0/r$ coordinate. Then 
\begin{equation}
v_f^2=\frac{ f(\rho)}{f_\infty}\Big[ 1 + \frac{m_1 \lambda \rho}{3 k^2} + \frac{m_1^2 \lambda^2 \rho^2}{9 k^4} - \Big(\frac{6 m_3 \lambda}{k^2}+\frac{2 m_1 m_2 \lambda^2}{k^4}+\frac{5 m_1^3 \lambda^3}{27k^6}\Big)\rho^3\Big] +\mathcal{O}(\rho^4)
\end{equation}
where $k=1-2\lambda f_\infty$. Near horizon the causality requires $m_1\lambda\leq 0$. In the case of $m_1=0$, the $m_3$ term in the bracket survives and gives $m_3\lambda>0$.

The final stage is the full theory including the quasi-topological term. In the large momentum limit, the dominant term is the last line of (\ref{casuality-eom}). This is the same as Einstein gravity, this time including $\lambda$ and $\mu$,
\begin{align}
v_f^2=\frac{ f(\rho)}{f_\infty}=& 1 + \frac{m_1  \rho}{3 f_\infty (1-2\lambda f_\infty - 3 \mu f_\infty^2)}  \nonumber\\
 & + \frac{m_1^2(\lambda+3\mu f_\infty) + 9 m_2 (1-2\lambda f_\infty - 3 \mu f_\infty^2)^2}{9 f_\infty (1-2\lambda f_\infty - 3 \mu f_\infty^2)^3}\rho^2  +\mathcal{O}(\rho^3)
\end{align}
Since $(1-2\lambda f_\infty - 3 \mu f_\infty^2)$ is proportional to the central charge, it is positive by the unitarity. Thus for small $\rho$,  $m_1$ should be non-positive. If we consider $m_1=0$, then
\begin{equation}
v_f^2=\frac{ f(\rho)}{f_\infty}= 1 + \frac{  m_2}{ f_\infty (1-2\lambda f_\infty - 3 \mu f_\infty^2)}\rho^2  +\mathcal{O}(\rho^3)
\end{equation}
leads to $m_2\leq 0$. We can go further and set $m_1=m_2=0$, then we need to expand the front velocity to order $\rho^3$,
\begin{equation}
v_f^2= 1 + \frac{2  m_3}{ f_\infty (1-2\lambda f_\infty - 3 \mu f_\infty^2)}\rho^3  +\mathcal{O}(\rho^4)
\end{equation}
This implies $m_3\leq 0$. In summary, causality indicates mass parameters to be nonpositive. This can be seen as $c_i\leq 0$ in the action (\ref{Action}). By the way, $m_i$'s are bounded below by requiring the temperature to be nonnegative, so from (\ref{temperature}),
\begin{equation}\label{nonnegative-temp}
1+M_1 \geq 0 \Rightarrow  m_1+2m_2+2m_3 \geq -4
\end{equation}

 \section{Conclusion}
The higher curvature gravities are important  in the holographic study of conformal field theories. In contrast to Einstein gravity which duals to four dimensional CFT's with equal central charges, higher curvature theories provide dual CFT's with two distinguished charges. A fundamental higher derivative gravity is expected to be derived from a string theory calculation, however the Gauss-Bonnet and cubic quasi-topological higher curvature gravities may be considered as toy models with rich structure to investigate the dual quantum field theory on the boundary \cite{Oliva:2010eb,Myers:2010ru,Myers:2010jv}. Of course, the existence of a dual theory to the Guass-Bonnet gravity is under question \cite{Camanho:2014apa,Cheung:2016wjt}, however, we merely considered it as a toy model. 

Here we studied a higher curvature massive gravity. The latter is important as generalization of the Einstein gravity and has phenomenological applications and theoretical consequences. It is known that when one considers massive gravity as a bulk theory in the context of AdS/CFT correspondence, the boundary theory violates the Lorentz invariance \cite{Vegh:2013sk,Davison:2013jba}. This is related to introducing the reference metric. Here we investigated the conformal structure of the boundary theory, if any. It was shown that adding a mass term to the bulk is equivalent to existence of a massive operator on the boundary. It appears like a short range Yukawa potential in the energy-momentum two-point function. The constant coefficient of this two-point function is proportional to the $c$ central charge and remains intact in the massive theory. Based on the null energy condition, we introduced a monotonically $a$-function which is a candidate for an $a$-theorem. 

In the second part of this article, we worked out an exact black brane solution. We derived the temperature to be mass dependent. Then we considered the hydrodynamic limit in the dual theory and calculated the shear viscosity to the entropy density, $\eta/s$ which is better to be interpreted as the rate of logarithm of the entropy production due to a strain. We found in (\ref{our}) that it includes two factors, an effective coupling due to higher curvature terms and an extra factor $\phi_0(r_0)^2$ from the massive gravity which we found it numerically. This later factor is less than one in the physical range of parameters which indicates that generally the graviton mass effect is reduction of the viscosity.

At the final stage, we investigated physical constraints as unitarity, ghost free and causality. The latter set condition on mass parameters $c_1$, $c_2$ and $c_3$. It was found that $c_1\leq 0$, if $c_1=0$ then $c_2\leq 0$ and so on. Of course, they are bounded from below due to non-negativity of temperature (\ref{nonnegative-temp}).

\vspace{1cm}
\noindent {\large {\bf Acknowledgment} }  Authors would like to thank A. Imaanpur and M. M. Sheikh-Jabbari for useful discussions.


\end{document}